\documentclass[12pt,a4paper]{article}

\usepackage[english]{babel}
\usepackage[utf8x]{inputenc}
\usepackage[T1]{fontenc}
\usepackage[a4paper,top=3cm,bottom=2cm,left=2.5cm,right=2.5cm,marginparwidth=1.cm]{geometry}

\usepackage{amsmath}
\usepackage{graphicx}
\usepackage{amssymb}
\usepackage[colorlinks=true, allcolors=blue]{hyperref}
\usepackage{authblk}
\usepackage{tikz} % for diagrams
\usetikzlibrary{arrows}
\usetikzlibrary{decorations.markings}
\usepackage{caption}
\usepackage{indentfirst} % indent at the beginning of each section
\usepackage{cite}
\usepackage{url}

\def\cm{{\,\rm cm}}
\def\m{{\,\rm m}}
\def\mb{{\,\rm mb}}
\def\km{{\,\rm km}}
\def\s{{\,\rm s}}
\def\eV{{\,\rm eV}}
\def\GeV{{\,\rm GeV}}
\def\TeV{{\,\rm TeV}}
\def\vd{{v_{\rm d}}} % drift velocity
\newcommand{\lr}[1]{\langle #1 \rangle}

\title{Strongly interacting dark matter \\ and the DAMA signal}
\author[]{Maxim Laletin}
\author[]{Jean-Ren\'{e} Cudell}
\affil[]{Institute of Space sciences and Technologies for Astrophysics Research, \\ Universit\'{e} de Li\`{e}ge, B\^{a}t B5A, Sart Tilman, 4000 Li\`{e}ge, Belgium}
\date{}

\begin{document}
\maketitle

\begin{abstract}
We show that models of strongly interacting (SIMP) dark matter built to reproduce the DAMA signal actually cannot account for its time dependence. We 
discuss the constraints on this type of models coming from direct detection experiments and study the propagation of thermalised dark matter 
particles in the ground for the allowed values of the parameters. We consider a simple 1D diffusion and a more detailed 3D diffusion. In both 
cases the predicted signal has either the wrong phase of the annual modulation or a much larger amplitude of the diurnal modulation.
\end{abstract}

\section{Introduction}
The DAMA and DAMA/LIBRA results have been with us for 20 years \cite{Bernabei:1998td}, and at present the observed modulation is the most 
significant unexplained experimental result in physics, reaching a level of 12.9 $\sigma$ \cite{Bernabei:2018corr}. The signal consists 
of hits in NaI scintillators, and corresponds to the emission of electromagnetic radiation with energies in the range 1 to 6 keV. To 
disentangle this signal from backgrounds, DAMA/LIBRA monitors it as a function of time. If the signal is due to weakly interacting massive particles (WIMP) scattering on the detector nuclei, and if dark matter is at rest with respect to our galaxy, then the motion of Earth in the galaxy creates a specific time 
signature, which is the signature. More specifically, the Sun moves towards the Cygnus constellation at roughly 220 km/s, and the Earth 
moves around the Sun at about 30 km/s. Depending on the time of the year, these velocities will add or subtract. As the Earth velocity 
is never exactly parallel or antiparallel to that of the Sun, only the projection of the velocity onto the Sun's velocity modulates the flux, 
and this amounts to a variation of about 6 \% of the flux of dark matter. What DAMA has been detecting for 20 years is such a 
periodic signal, and it has precisely the correct phase, being maximum around June 2nd and minimum around December 2nd.
 Besides this modulation, the bulk of the dark-matter flux and the residual backgrounds should also produce events.
 
This could have been the answer to the 100-year old puzzle of dark matter, but the rub has been that no other detector sees a signal,
and several of them are severely incompatible with the DAMA/LIBRA results\footnote{Some studies indicate that the new DAMA data \cite{Bernabei:2018corr} is itself inconsistent with the conventional isospin-conserving spin-independent WIMP interaction with nuclei \cite{Baum:2018ekm}.} \cite{Billard:2013qya,Aprile:2017iyp,Akerib:2013tjd}.
A number of explanations of this problem have been tried, assuming that the DAMA signal is due to conventional physics, but all of them 
have been rebutted \cite{Bernabei:2018corr,Bernabei:2014tqa}. As no standard physics explanation seemed to work, several groups
proposed models that could lead to a signal in DAMA and not in other detectors. We are aware of three classes of such models:
mirror matter \cite{Addazi:2015cua,Cerulli:2017jzz}, resonant dark matter \cite{Bai:2009cd}, and specific strongly interacting massive particle (SIMP) models \cite{Wallemacq:2013hsa,Wallemacq:2014sta}.
In this paper, we shall concentrate on the latter.

Models based on SIMPs find their origin in the work of M. Khlopov and collaborators, who proposed that a composite bound state, made of a 
SIMP of charge $-2$ and a $^4$He nucleus, could be the dominant form of dark matter \cite{Fargion:2005ep} and the key to the explanation of 
the DAMA signal \cite{Khlopov:2010ik,Khlopov:2011tn}. The elastic cross section of this OHe atom is comparable to that of neutrons, and after
hitting the ground it will quickly thermalise and turn into a cloud of slowly-moving heavy particles. As it is moving with thermal energy, it does not 
create detectable recoils when it hits nuclei. Instead, it was argued \cite{Khlopov:2010ik} that a repulsive force would arise when OHe got
close to a nucleus, and that it might allow bound states with sodium. The formation of these bound states would result in the emission of
photons of a few keV energy, which would make the DAMA signal. The demise of this model came from the absence of such a repulsive force 
\cite{Cudell:2012fw}, which would have observable and disastrous consequences \cite{Cudell:2014dva}.

But the essential features of this model could be preserved, in the context of a dark sector, with a subdominant component made of dark atoms
\cite{Wallemacq:2013hsa} or dark antiatoms \cite{Wallemacq:2014sta}. The elastic cross sections could be made large enough for the incoming 
composite SIMPs to thermalise and escape detection from nuclear recoil. It was also possible to produce a signal via binding with some of the 
elements of the DAMA detector, and not with those of other detectors. These two models are 
 examples of a class of models which could, in
principle, reproduce the DAMA signal. However, as we shall show in this paper, the first ingredient, i.e. the thermalisation of the SIMPs before 
they reach the detector, makes it impossible to reproduce the time signature observed by DAMA \cite{Bernabei:2014jnz}.

This paper is organised as follows: in Section \hyperref[constr]{2}, we spell out the constraints on the elastic cross section of SIMPs, in Section \hyperref[prop_models]{3} we explain the 
various propagation models that we used, and show the time dependence of the signal according to the degree of precision of the diffusion 
model, in Section \hyperref[grav_focus]{4} we discuss the impact of the gravitational focusing of the Sun on the signal. We then conclude about the consequences of this work for SIMPs and composite dark matter.

\section{Constraints}
\label{constr}

\begin{figure}[t!]
	\center{\includegraphics[width=1\textwidth]{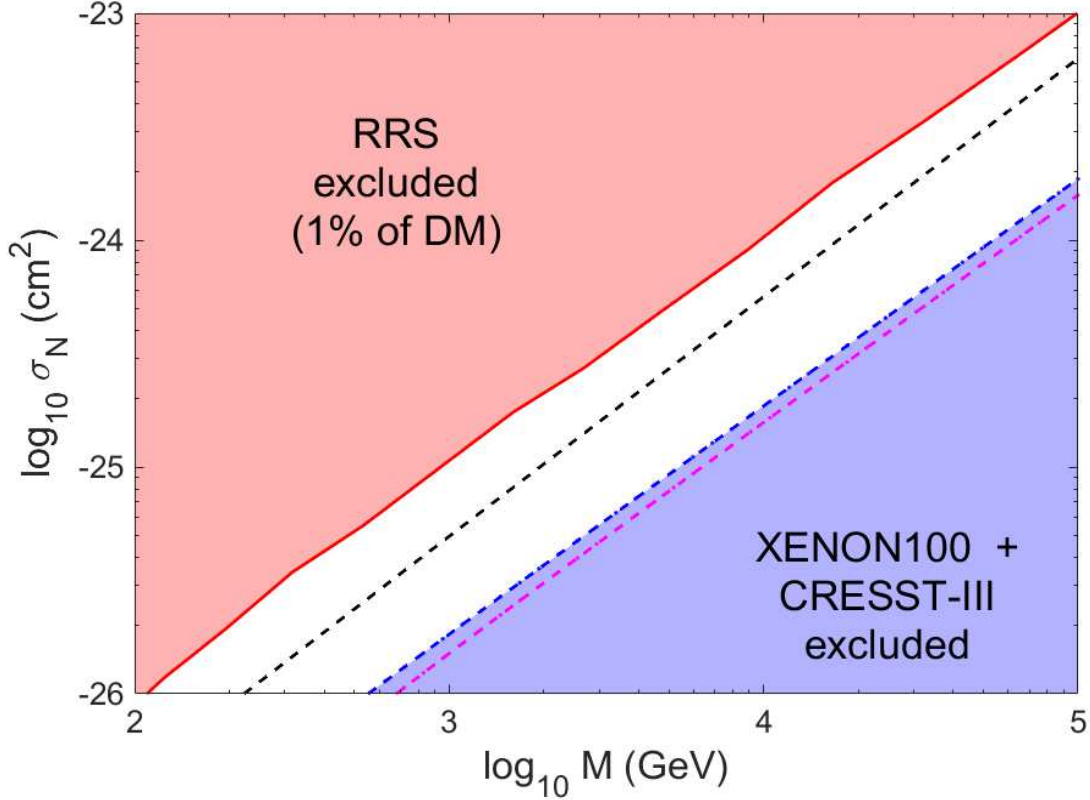}}
	\caption{Exclusion limits on the SIMP-nucleus elastic scattering cross section. The area above the red curve is excluded by the RRS 
	experiment \cite{Rich:1987st} for SIMPs making up 1\% of the local dark-matter density. The area below the dot-dashed curve is excluded by the 
	CRESST-III \cite{Petricca:2017zdp} (blue curve) and the XENON100 \cite{Aprile:2016wwo} (magenta curve) data. The black dashed curve 
	indicate the points of the parameter space which correspond to the value of thermalisation depth equal to the depth of the LNGS (1.4 km).}
	\label{constraints} 
\end{figure}
As the problem will come from thermalisation, we can study a generic class of models. We do not need to discuss the particular mechanism which 
creates the signal in the detector and we can consider only the parameters that determine the propagation, i.e. the particle mass $M$ and the 
cross section $\sigma_{N}$ of elastic scattering on the nuclei of the ground.
The depth of thermalisation is given by
\begin{equation}
l_{\rm th} = \left( \frac{M}{m} \right) \log \left( \frac{v_0}{v_{\rm th}}  \right) \frac{1}{n\sigma_N} ,
\label{l_th}
\end{equation}
with $m$ the mass of atoms in the crust, $v_0$ the incoming velocity, $v_{th}$ the thermal velocity, and $n$ the 
density of the crust. For example, to thermalise at the depth of LNGS (1.4 km) a particle with a mass of 1 TeV should have $\sigma \approx 5 \ 10^{-26} 
\cm^2$. 
The first set of constraints, shown in Fig.~\ref{constraints}, actually comes from underground direct-detection experiments with null results. The 
experiments that are the most sensitive\footnote{This is due to a combination of multiple factors, such as the threshold energy, detector material 
and the depth of the laboratory.} to the particles we consider are XENON100 ($E_{\rm th} = 700\eV $) \cite{Aprile:2016wwo} and CRESST-III 
($E_{\rm th} \approx 100\eV $) \cite{Petricca:2017zdp}, located at the same site as DAMA.

Experiments operating at the top of the Earth atmosphere or above provide the second set of constraint on SIMPs, and limit the parameter space 
from above. The most severe constraint comes from the balloon-borne direct detection experiment RRS \cite{Rich:1987st}. For example, the 
aforementioned cross section of $\sim 10^{-26} \cm^2$ lies a few orders of magnitude above the limit for 1 TeV particle. To loosen this constraint 
one has to assume that SIMPs are a subdominant fraction of the local dark-matter density. Taking into account the first set of constraints this fraction has 
to be $< 5\%$. Hereafter we take the fraction of SIMPs to be 1\%, although we shall also study one particular case where this fraction is $0.1 \%$. 
Note, that these constraints are for the elastic scattering cross section on silicon nuclei and were taken from the analysis of the RRS data 
\cite{Rich:1987st}, not recalculated from the constraints on dark-matter-nucleon cross section (e.g., as in \cite{Erickcek:2007jv}). Since it is a reasonable 
assumption to take the average rock as composed entirely of silicon atoms (see below), we can apply these constraints directly.

\section{Propagation models}
\label{prop_models}

\subsection{Ray propagation}
\label{instant_prop}
\begin{figure}[h]
	\center{\includegraphics[width=0.6\textwidth]{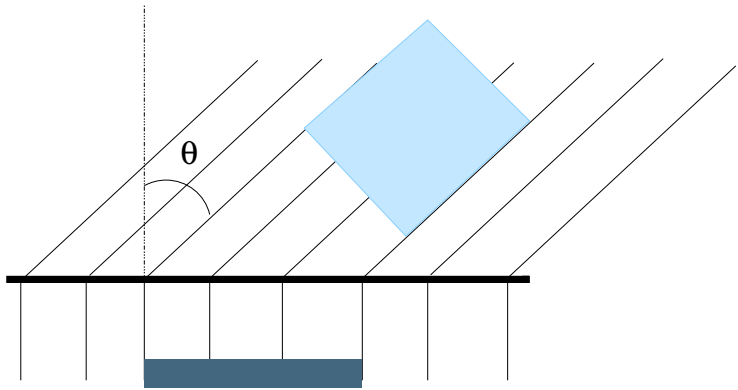}}
	\caption{The geometry of a wave packet hitting the ground.}\label{rays}
	\label{ray_prop} 
\end{figure}

To explain the problem, let us first use a very simplified picture of propagation. Assume that the ground can be treated as an infinite plane, on 
which an infinite beam of dark matter falls, and assume that all the particles have the same velocity $V_{\rm DM}=v_0$. This beam then hits the ground,
the particles slow down, and after a while move down because of gravity. Consider a packet of SIMPs travelling through the ground, as in Fig.~\ref{ray_prop}. 
As the speed is lower below ground, the packet gets concentrated. But the packet also gets spread, as its section is inversely proportional to 
$\cos\theta$, with $\theta$ the angle with the azimuth, and becomes zero if $|\theta|>\pi/2$, i.e. when the flux of dark matter comes from below 
the horizon. So the density of the packet is proportional to $\cos\theta$. $\theta$ is a complicated function of time which can be found in 
\cite{Bernabei:2014jnz}. This function depends on the Earth rotation, and the flux of dark matter is screened by the Earth part of the day for a 
period of the year. This leads to a daily variation, and to a phase shift in the yearly flux.

\begin{figure}[t]
	\begin{minipage}[h]{0.5\linewidth}
		\center{\includegraphics[width=1\textwidth]{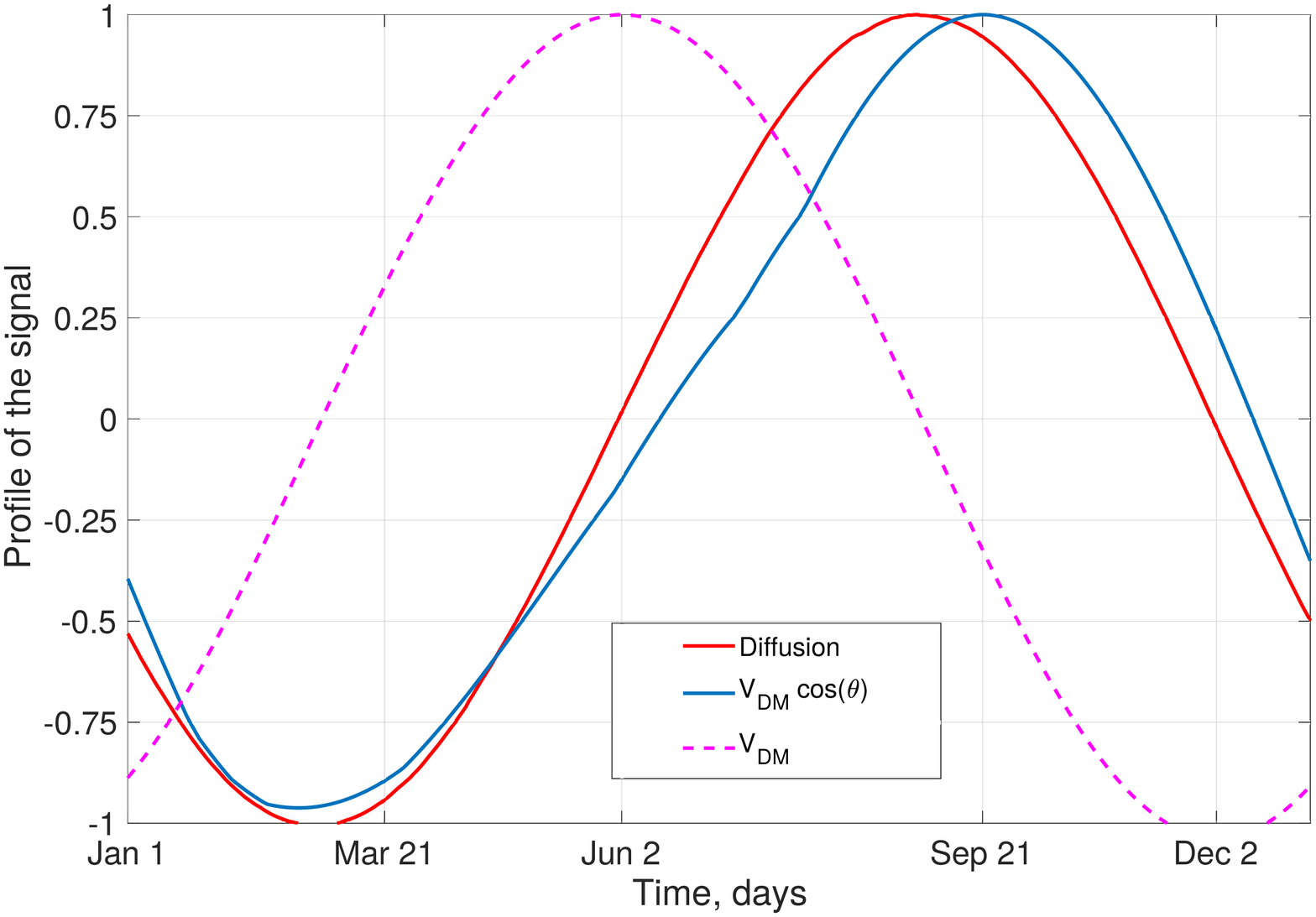}}
	\end{minipage}
	\begin{minipage}[h]{0.5\linewidth}
		\center{\includegraphics[width=1\textwidth]{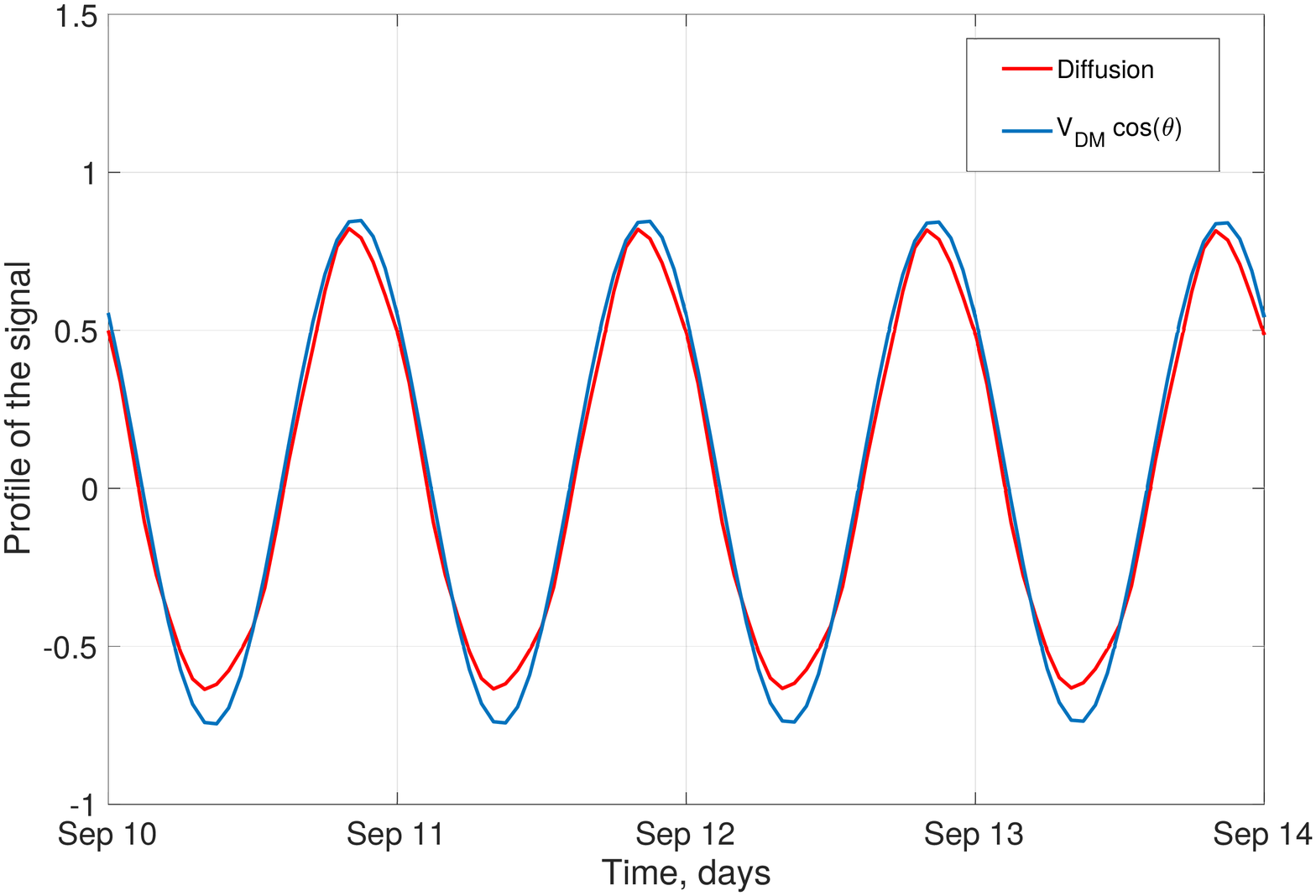}}
	\end{minipage}
	\caption{The time profiles of the normalized residual rate of the signal in DAMA. The \textit{left} plot shows the daily-averaged annual cycle of 
	the rate, proportional to $V_{\rm DM}$ (WIMPs, dashed magenta curve), $V \cos{\theta}$ (ray propagation of SIMPs, blue curve) and the one obtained 
	within the diffusion propagation model for SIMPs ($M = 1 \TeV$ and $\sigma = 60 \mb$, red curve). The \textit{right} plot shows the 
	corresponding daily modulation of the signal rate over a period of 5 days for the ray propagation model (blue curve) and for the diffusion 
	propagation model (red curve).}
	\label{cosine_vs_diffusion}
\end{figure}

The resulting time profile of the expected signal in the DAMA detector is shown in Fig.~\ref{cosine_vs_diffusion}, where we compare it to the 
WIMP signal and to the results of the diffusion approximation, which we shall discuss later. One immediately notices a substantial time delay of 
the annual modulation (left plot) compared to the WIMP case that reproduces the DAMA data \cite{Bernabei:2013xsa}, as well as a non negligible
diurnal modulation of the signal (right plot). A modulation of the daily signal is actually present in the WIMP case as well, because of the change of flux due to the Earth rotation, but its amplitude is tiny.

We treat this result as an example of the difficulties that a more detailed model of propagation has to overcome in order to provide a better fit of 
the DAMA data. One could hope that diffusion will spread the daily variation enough to make it unobservable. The annual phase shift could 
also be affected by a more precise treatment, which we shall consider now.

\subsection{Diffusion}
\label{diffusion_prop}

A more realistic treatment of the propagation of thermalised SIMPs in the ground has to account for the fact that it takes time for these particles to 
reach the detector from the point where they are in equilibrium with the surrounding matter. Furthermore, they constantly scatter on nuclei in the 
medium and get driven by the gravitational field, so they do not proceed in the same direction as the incoming flux of dark matter. In principle, a fraction of the incoming flux can even scatter back up, out of the ground. We shall assume that once a SIMP gets above the ground it is 
never coming back (at least, in the same vicinity), so particles, which thermalise close to the surface have lower chances to contribute to the 
signal.

The processes, described above, resemble very much the diffusion of a gas in the presence of a gravitational field, so we are going to adopt the 
corresponding diffusion equation for the number density of SIMPs as a function of space and time $N(\vec{x},t)$ to see how it evolves inside the 
detector: 
\begin{equation}
\frac{\partial N(\vec{x},t)}{\partial t} = D\Delta N(\vec{x},t) - \vd\frac{\partial N(\vec{x},t)}{\partial z} + f(\vec{x},t) \,,
\label{diffusion}
\end{equation}
where the first term on the right-hand side accounts for diffusion, with $D$ the diffusion coefficient, the second term describes the influence of gravity, with $\vd$ the 
drift velocity (parallel to the z-axis) 
and the third term $f(\vec{x},t)$ is the source function. 
One should, in principle, solve Eq.~\eqref{diffusion} in an 
inhomogeneous unbounded space, as the air and the ground have different diffusion coefficients, for a source function steady in time\footnote{The solution of the heat transfer equation in a medium, 
consisting of regions with different values of 
thermal conductivity, was found by Sommerfeld in \cite{Sommerfeld1894}.}. However, the density of 
air is much smaller than that of rock, so we can neglect the propagation in the atmosphere and solve the diffusion equation in the 
space limited by the surface $S_{\rm b}(x,y)$ between the two media with the condition $ \left. N(\vec{x},t) \right|_{S_{\rm b}} = 0 $ imposed on 
this boundary. Physically, this conditions corresponds to a sink of particles at the boundary.

The diffusion coefficient depends on several parameters: those characterizing the medium, the mass of the SIMPs and the cross section 
describing their interaction with nuclei in the medium. We show in 
Appendix~\ref{diffusion_derivation} that the following formula holds:
\begin{equation}
% D = \frac{m+M}{mn\sigma_{\rm N}}\sqrt{\frac{\pi kT}{8 M}} = \frac{\pi}{8}\left(\frac{m+M}{m}\right)\lambda v_{\rm th}.
D = \frac{\pi}{8} \frac{M}{\mu} \lambda \, v_{\rm th} = \frac{m+M}{mn\sigma_{\rm N}}\sqrt{\frac{\pi kT}{8 M}} \, ,
\label{diff_coeff}
\end{equation}
where $\mu = mM/(m+M)$ is the reduced mass, $n$ is the number density of the medium, $\lambda = (n\sigma_{\rm N})^{-1}$ is the mean free 
path and $v_{\rm th}$ is the 
average thermal velocity of SIMPs. Using the Einstein relation \cite{1905AnP322549E} $D = \alpha kT$, where $\alpha$ is the ratio of the drift 
velocity $\vd$ to the 
applied force $F = Mg$ with $g=9.81$ m/s$^2$ the gravitational acceleration, one gets the expression of the drift velocity in the gravitational field
\begin{equation}
\vd = \frac{Mg\lambda}{\mu v_{\rm th}}\, .
\label{drift_vel}
\end{equation}
For simplicity, we assume that the rock is composed only of silicon atoms with $m = 28 \GeV$, following \cite{Sigurdson:2004zp}. Although 
oxygen 
atoms are actually about 4 times more abundant in rock than silicon atoms, we have checked that our results are insensitive to this, and that our 
main conclusions remain the same.

The source function $f(\vec{x},t)$ describes the rate of the thermalised SIMPs density increment and depends on many factors. First of all, it 
depends on the local number density of SIMPs, which we take to be $n_{\rm loc} = 0.01 \cdot \rho_{\rm loc} / M$, as explained in Sec.~\ref{constr}, where $\rho_{\rm loc} = 0.39 \GeV / \cm^3$ is the conventional value of the local dark-matter density. Second, SIMPs are coming with 
different velocities, which follow the velocity distribution $\omega_{v}$, hence the distance they travel through the ground until thermalisation 
also differs. Let $\vec{l}$ be the vector connecting the point $(x^*,y^*)$ on the surface $S_{\rm b}$, where the particle enters the ground, with the 
point $(x,y)$, where it acquires the thermal velocity. It is convenient to transform the velocity distribution into the path length distribution 
$\omega_{v} \rightarrow \omega_{l}$ (see Appendix \ref{append_distr}). Third, the rate of SIMPs crossing the surface $S_{\rm b}$ at the point $\left\lbrace x^*,y^*\right\rbrace$ depends on the angle between the velocity of the incoming beam of SIMPs and the normal vector to the surface at the given point $\vec{n}$. 

The source function can be constructed in the following way
\begin{equation}
f(\vec{x},t) = n_{\rm loc} \iint dx^*dy^* \; \omega_l(\vec{x^*} - \vec{x},t) \, v\left(\left|\vec{x^*} - \vec{x}\right|\right) \, .
\label{source}
\end{equation}

The complexity of the solution of Eq.~\eqref{diffusion} depends mainly on the shape of the boundary surface $S_{b}$. We are going to consider 
two relevant cases.

\subsubsection{The plateau approximation}

The simplest shape of the boundary surface $S_b$ is obviously a plane, for example $z = 0$. Since the total incident flux of SIMPs through the 
boundary is uniformly distributed on that surface the diffusion is symmetric in $x$ and $y$ and the density gradient is directed along the $z$-axis. 
In this case Eq.~\eqref{diffusion} becomes one-dimensional and its solution has the following form

\begin{equation}
N(z,t) = \int_{-\infty}^{t} d\tau \int_{0}^{\infty} d\xi \exp \left(\frac{2\vd(z-\xi)(t-\tau) - \vd^2 (t-\tau)^2}{4D(t-\tau)}\right) \; G(z - \xi,t-\tau) \; f(\xi,\tau) \,.
\label{solution_flat}
\end{equation}
Here $G(z - \xi,t-\tau)$ is the Green's function of the one-dimensional diffusion equation in the semi-infinite space with the boundary condition 
$N(0,t) = 0$ \cite[p.~209]{tikhonov1964}

\begin{equation}
G(z - \xi,t-\tau) = \frac{1}{\sqrt{4\pi D(t-\tau)}}\left[ \exp\left(- \frac{(z-\xi)^2}{4D(t-\tau)} \right) - \exp\left( -\frac{(z+\xi)^2}{4D(t-\tau)} \right)  \right] \,,
\label{greens_flat}
\end{equation}
which describes the propagation of particles created at the point $\xi$ at the moment $\tau$. The income of thermalised SIMPs is determined by 
the source function \eqref{source}, which also becomes effectively one-dimensional. The factor in front of the Green's function in 
Eq.~\eqref{solution_flat} appears due to the presence of the drift term in the Eq.~\eqref{diffusion} (see Appendix \ref{append_drift}). 

% \begin{equation}
% f(\xi,\tau) = n_{\rm loc} \iint dx^*dy^* \; \omega_l(\xi,\tau) \, v(l)
% \label{source_flat}
% \end{equation}

\begin{figure}[t]
\begin{minipage}[h]{0.52\linewidth}
\center{\includegraphics[width=1\textwidth]{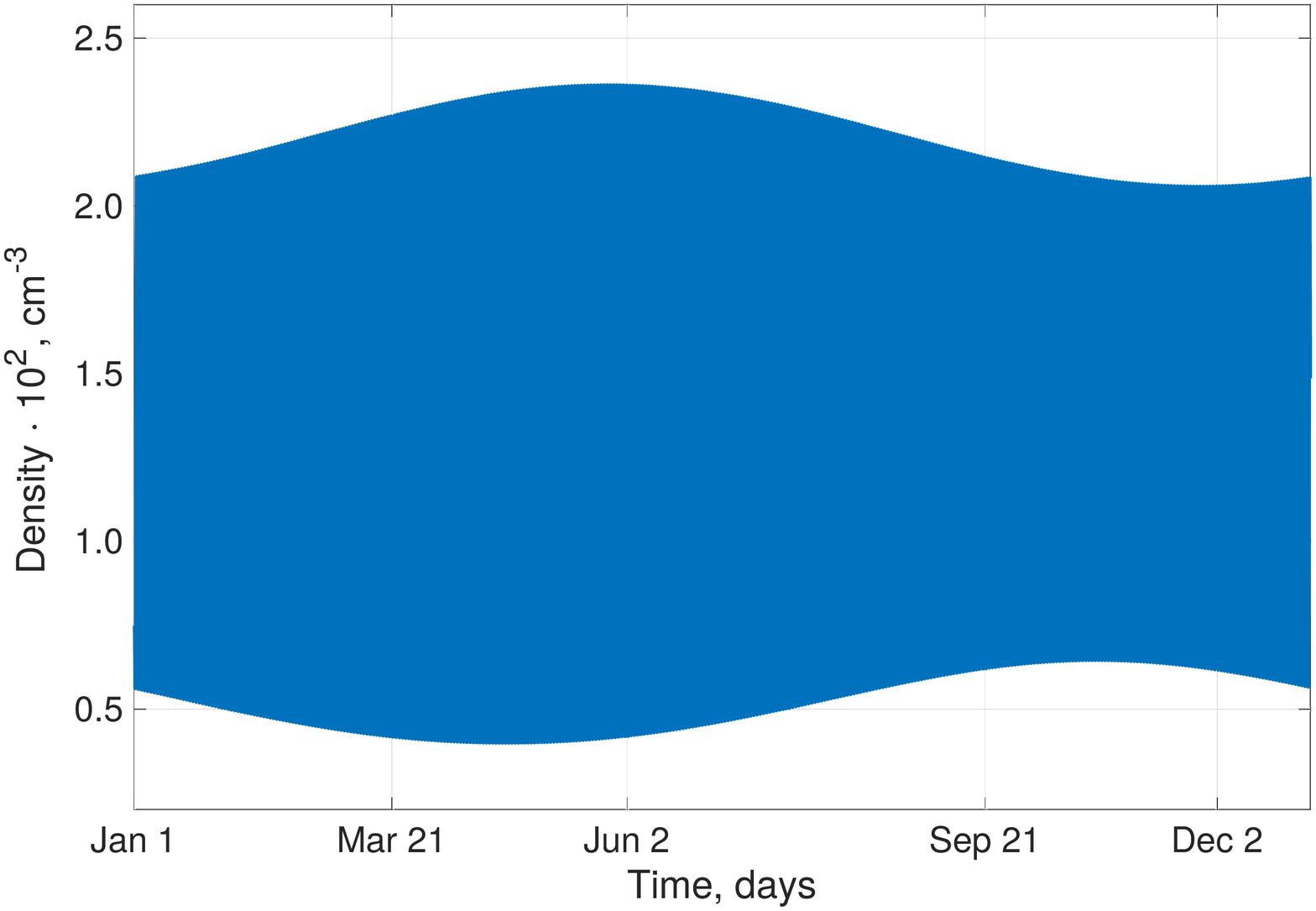}}
\end{minipage}
\begin{minipage}[h]{0.52\linewidth}
\center{\includegraphics[width=1\textwidth]{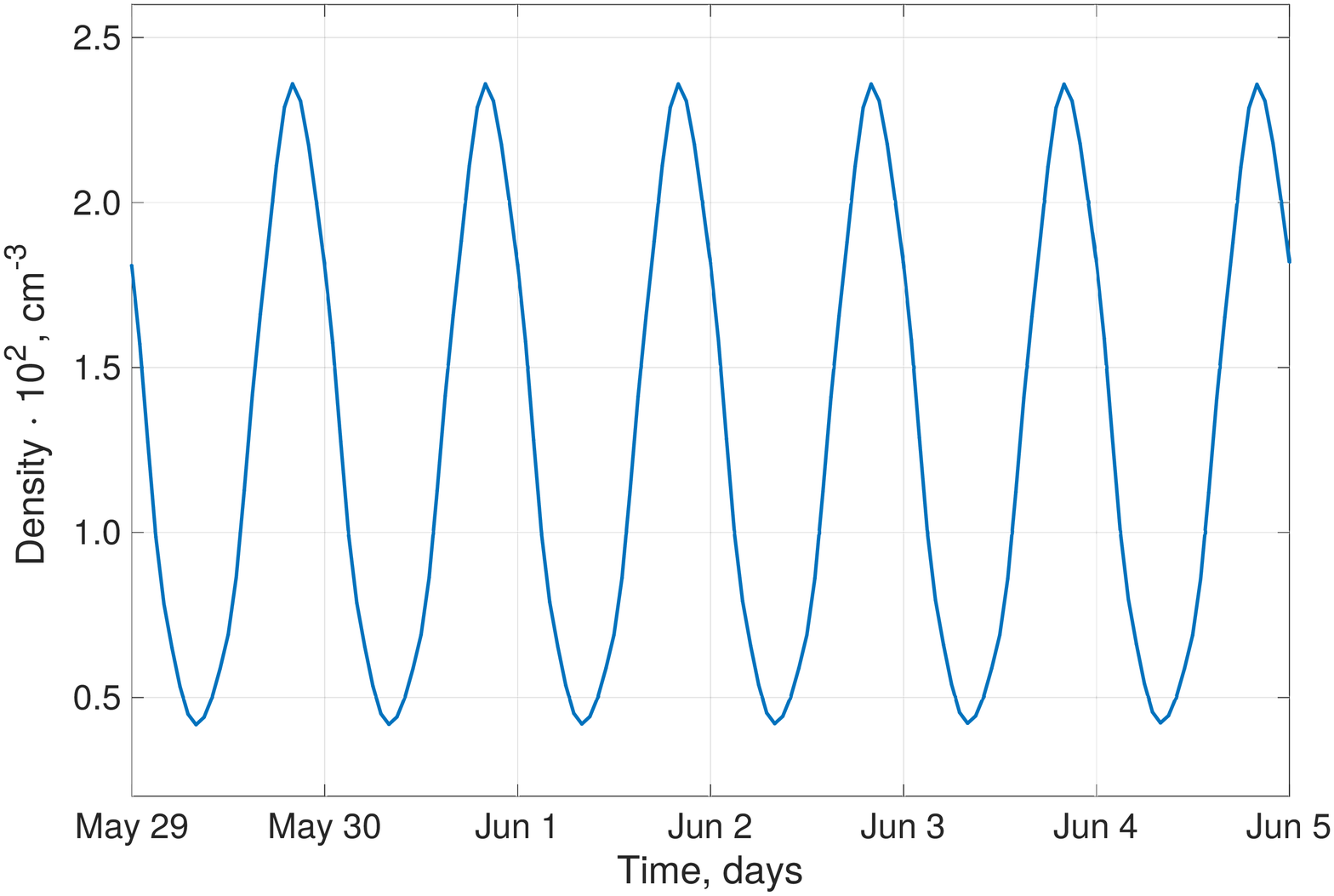}}
\end{minipage}
\caption{The time-dependence of particle number density of SIMPs with $M = 1 \TeV$ and $\sigma = 60 \mb$ inside the DAMA detector over the 
year (\textit{left}) and zoomed-in around June the 2nd (\textit{right}).}
\label{density_plot}
\end{figure}

The integration in Eq.~\eqref{solution_flat} can be performed numerically. As an example, we demonstrate the time dependence of the particle 
density in the DAMA detector ($z = 1.4 \km$) for SIMPs with $M = 1 \TeV$ and $\sigma = 60 \mb$ (Fig.~\ref{density_plot}). We assume that the signal in DAMA is 
proportional to the density of SIMPs in the detector $S = \beta N$, where $\beta$ is an additional parameter, which depends on the underlying 
physics of SIMP-nucleus inelastic interactions. In our analysis we only use this parameter to fit the predicted signal to the annual modulation data 
and we do not impose any constraints on its value. Following the procedure described in \cite{Bernabei:2014jnz} we calculate the residual rate of 
events and compare it to the DAMA data on annual and diurnal modulation %\cite{Bernabei:2014jnz}
in $2 - 6$ keV energy interval. 

\begin{figure}[!t]
	\begin{minipage}[h]{0.52\linewidth}
		\center{\includegraphics[width=1\textwidth]{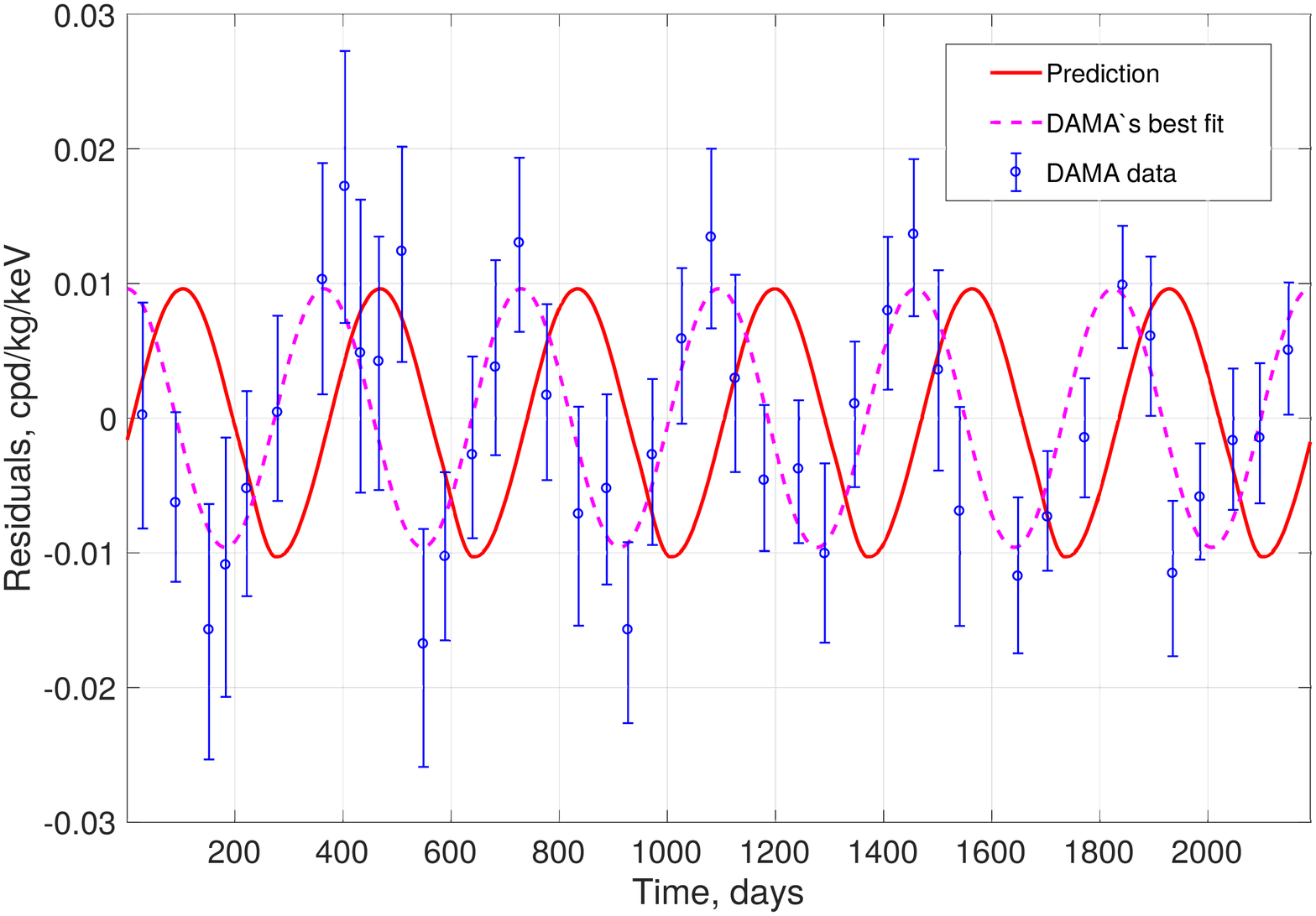}}
	\end{minipage}
	\begin{minipage}[h]{0.52\linewidth}
		\center{\includegraphics[width=1\textwidth]{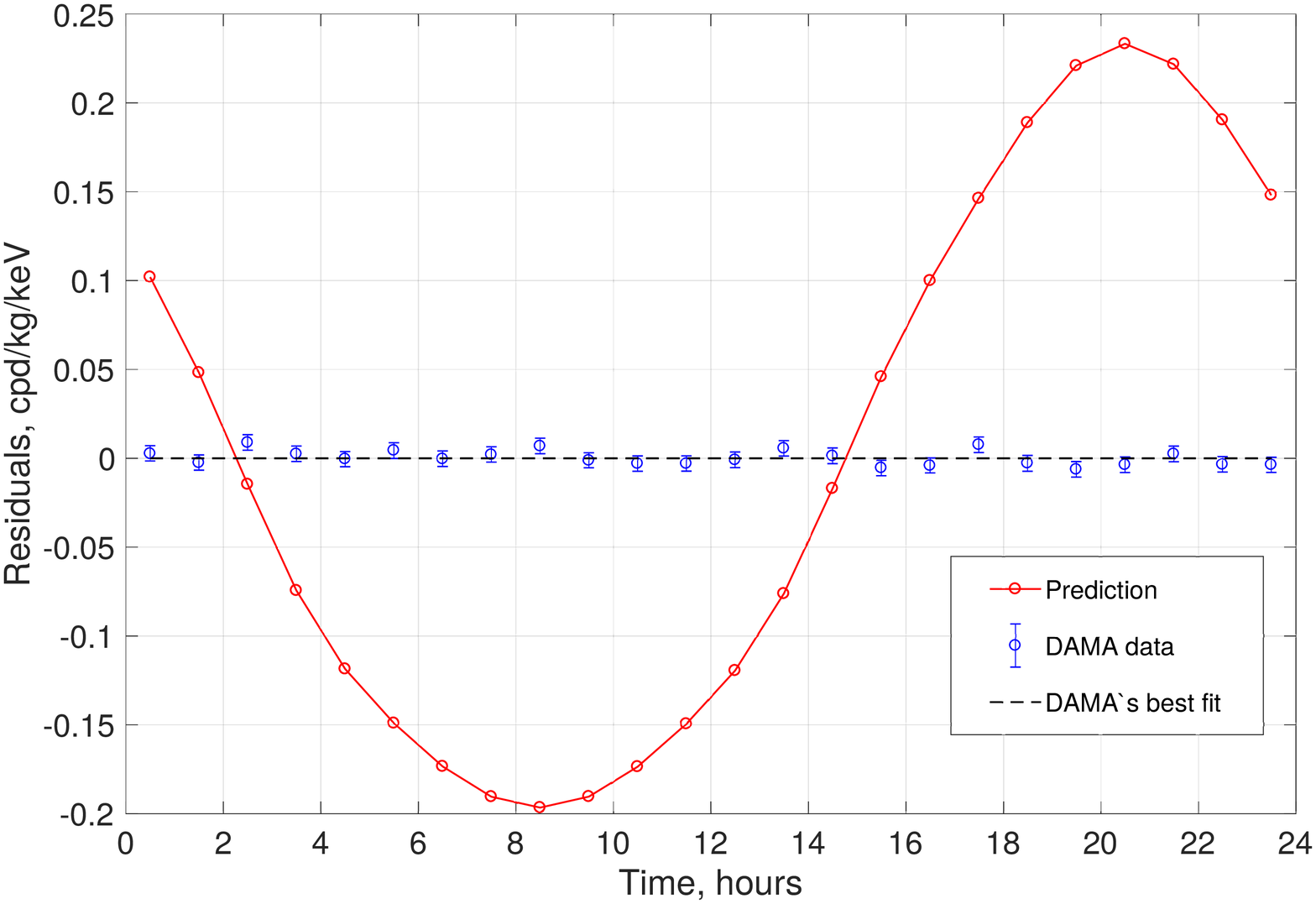}}
	\end{minipage}
	\caption{Annual (\textit{left}) and diurnal (\textit{right}) modulations of the residual rate of events in the DAMA detector in $2 - 6$ keV energy interval for SIMPs with $M = 1 \TeV$ and $\sigma = 60 \mb$ calculated in the plateau approximation and compared to the experimental data. Here and in the similar plots below the dashed curves indicate the best-fit model, which corresponds to the case of WIMPs.}
	\label{modulation_flat_plot}
\end{figure}

Fig. \ref{modulation_flat_plot} shows that for $M = 1 \TeV$ and $\sigma = 60 \mb$ not only the amplitude of the diurnal modulation of the signal 
exceeds the data points by orders of magnitude, but also the phase of the annual modulation of the signal is off the best-fit value by about
100 days. The daily variation comes from the fact that the density of dark matter hitting the ground depends on the angle between the flux of dark 
matter and the ground. The phase of the yearly signal comes from a change in the daily average of that angle due to the obliquity of the ecliptic.
Diffusion of dark matter is not strong enough to flatten the first variation, and it cannot change the second. This is mainly because the value of the 
diffusion coefficient allowed by the constraints in 
Fig.~\ref{constraints} is quite large $D \gtrsim 1000 \m^2/\s$, which corresponds to $\vd \sim 10 \m/\s$. Thus, it takes a cloud of SIMPs around a 
few minutes to get from the surface to the laboratory, which is totally negligible for the relevant observational time scales. 

\begin{figure}[!t]
	\begin{minipage}[h]{0.52\linewidth}
		\center{\includegraphics[width=1\textwidth]{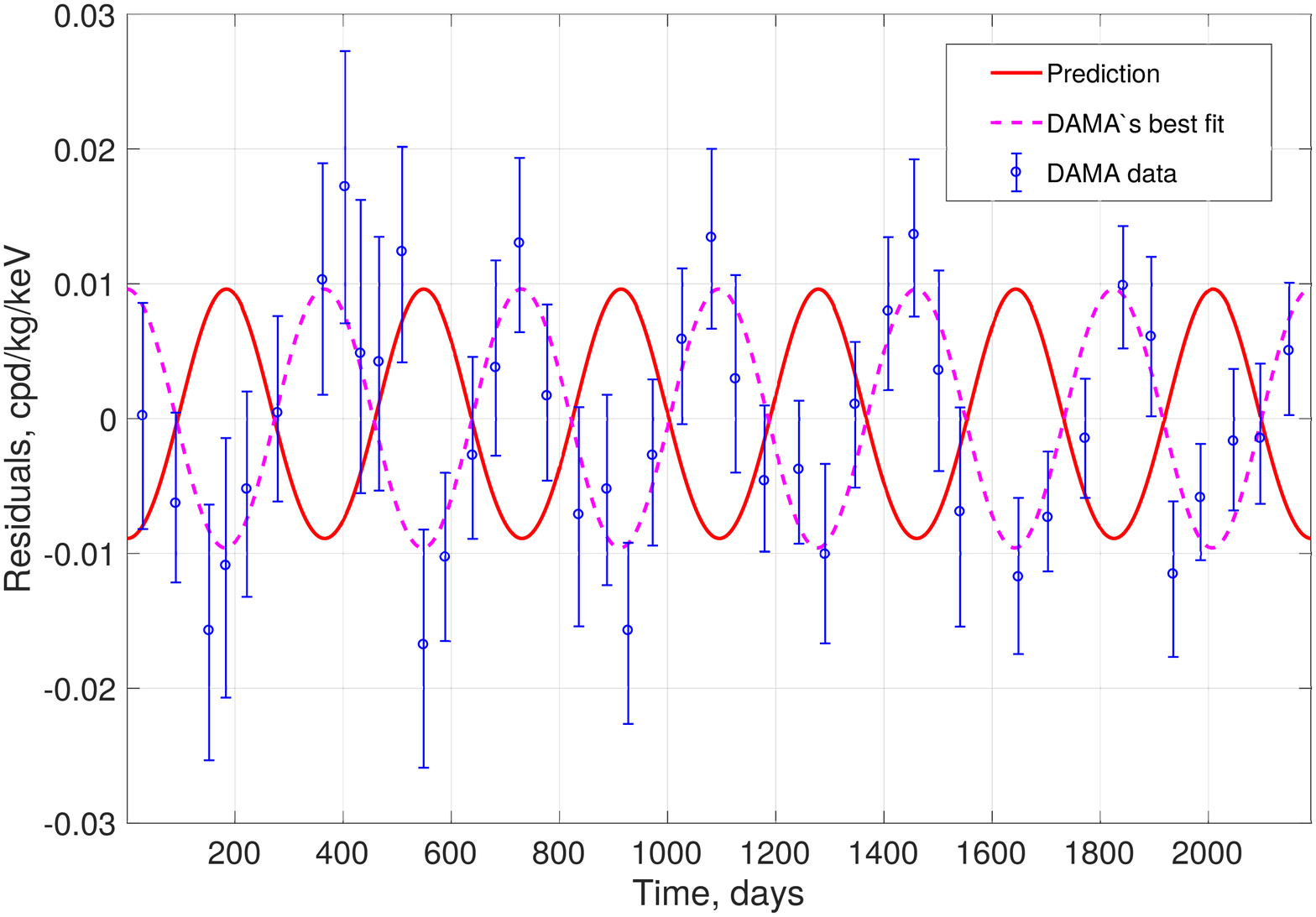}}
	\end{minipage}
	\begin{minipage}[h]{0.52\linewidth}
		\center{\includegraphics[width=1\textwidth]{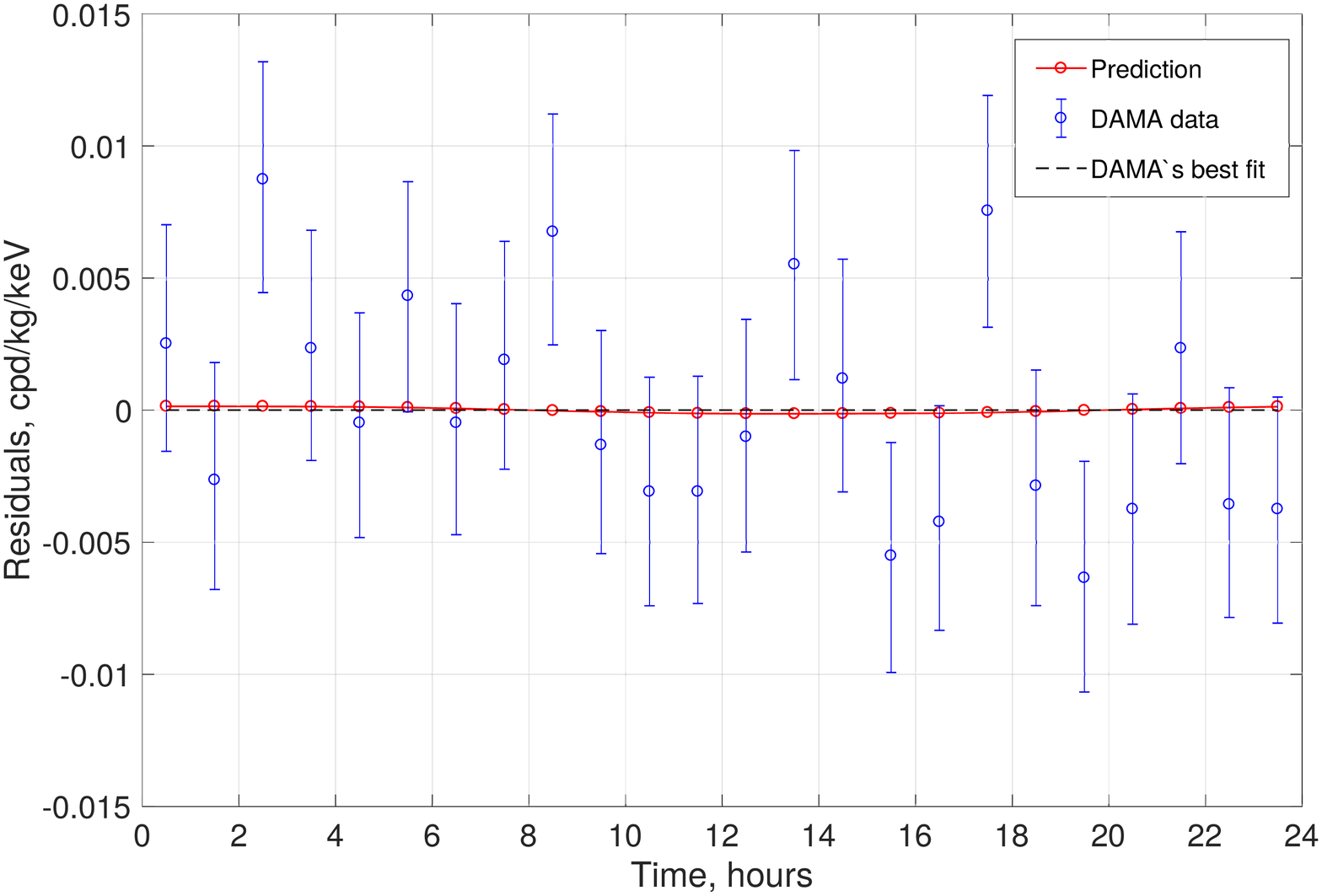}}
	\end{minipage}
	\caption{Annual (\textit{left}) and diurnal (\textit{right}) modulations of the residual rate of events in the DAMA detector in $2 - 6$ keV energy interval for SIMPs with $M = 1 \TeV$ and $\sigma =  5 \mb$ calculated in the plateau approximation and compared to the experimental data.}
	\label{modulation_flat_plot_deep}
\end{figure}

Qualitatively similar results are obtained for the whole allowed range of parameters above the black dashed line in Fig. \ref{constraints}. Returning 
to Fig.~\ref{cosine_vs_diffusion}, we see that the profile of the signal, predicted within the diffusion approach, does not substantially differ from the 
much simpler model that we considered before.

As the cross section gets smaller and the thermalisation depth distribution extends much deeper than DAMA, the diurnal modulation of the signal 
almost vanishes, though the phase of the annual modulation is still wrong (see Fig. \ref{modulation_flat_plot_deep}). Furthermore, the average 
density of SIMPs in the detector in this case is many orders of magnitude smaller than for $\sigma = 60 \mb$. Hence, to account for the observed 
rate, some extreme values of the parameter $\beta$ are required, which, in its turn, implies unrealistically high values of the inelastic cross section. Reducing the fraction of SIMPs to $0.1\%$ to lift the RRS constraint and allow larger cross sections neither recovers the right phase of the annual modulation, nor eliminates the diurnal modulation.

\subsubsection{Realistic Gran Sasso terrain surface}

\begin{figure}[!h]
\center{\includegraphics[width=1\textwidth]{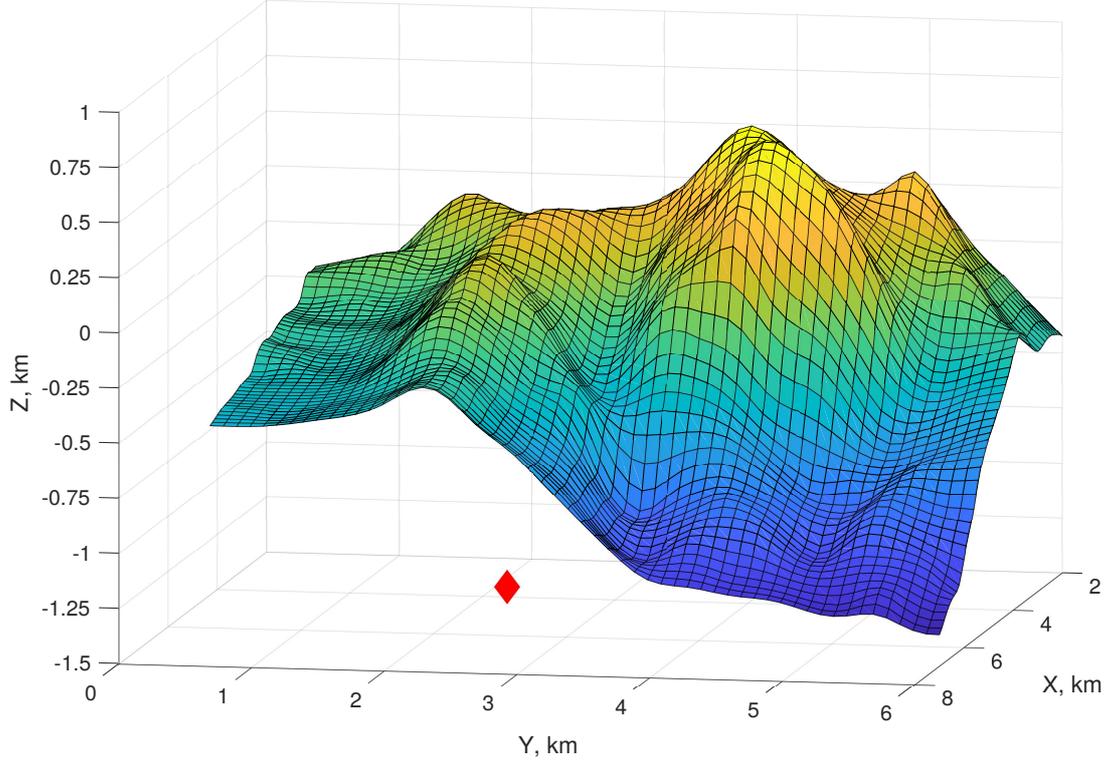}}
\caption{The reconstruction of the Earth's surface in the vicinity of LNGS. The location of the laboratory is marked with the red diamond. The elevation data is obtained with \cite{Google:elevation}. 
%\todo[inline]{preliminary picture}
}
\label{Gran_Sasso_surface}
\end{figure}

Clearly, the surface of the Earth around LNGS is far from being a simple plane. Finding the exact analytical solution (especially, the Green's 
function) of the diffusion equation \eqref{diffusion} for a realistic humpy terrain as the one surrounding LNGS (see Fig. \ref{Gran_Sasso_surface}) 
is very complicated. The problem can be drastically simplified if one neglects the loss of particles through the boundary into the 
atmosphere\footnote{We have studied this simplification in the plateau geometry and our results indicate that the difference between the 
exact solution and the simplified one is not larger than $\approx 10\%$.} and uses a simple Green's function of the 3D diffusion in the unbounded 
space 
\begin{equation}
G(x - \chi, y - \upsilon, z - \xi,t-\tau) = \frac{1}{(4\pi D(t-\tau))^{3/2}} \exp\left( \frac{(x-\chi)^2 + (y-\upsilon)^2 + (z-\xi)^2}{4D(t-\tau)} \right).
\label{greens_unbound}
\end{equation}
% \begin{equation}
% N(\vec{x},t) = \int_{0}^{t} \iiint_{-\infty}^{\infty} d\vec{\xi} \exp \left(\frac{2\vd(z-\xi_z) - \vd^2 (t-\tau)^2}{4D(t-\tau)}\right) \; G(\vec{\xi},t-\tau) \; f(\vec{\xi},\tau) \, , 
% \end{equation}
To further simplify the calculation we break the considered landscape into squares of the size $S = 500 \times 500 {\rm \, m}^2$ and construct the 
source function piecewise
\begin{equation}
f(\chi,\upsilon,\xi,t) \approx n_{\rm loc} \, S \sum_i \sum_j \; \omega_l(x_i - \chi,y_j - \upsilon,z_{ij} - \xi,t) \, v(l)\,,
\label{source_piece}
\end{equation}
where $x_i$ and $y_j$ are the coordinates of the centre of each square and $z_{ij}$ is their elevation. The density of SIMPs at the given point in 
time and space can be calculated as follows\footnote{For the sake of brevity, in this expression we do not explicitly indicate that the source 
distribution is truncated above the boundary surface.}
\begin{multline}
N(x,y,z,t) \approx n_{\rm loc} \, S \sum_i \sum_j \int_{-\infty}^{t} d\tau \int_{-\infty}^{\infty} d\chi \int_{-\infty}^{\infty} d\upsilon \int_{0}^{\infty} d\xi \; 
\times \\
\times \; G(x - \chi, y - \upsilon, z - \xi,t-\tau) \, \exp \left(\frac{2\vd(z-\xi)(t-\tau) - \vd^2 (t-\tau)^2}{4D(t-\tau)}\right) \; \times \\
\times \omega_l(x_i - \chi,y_j - \upsilon,z_{ij} - \xi,t) \, v(l) \,.
\label{solution_realistic}
\end{multline}

\begin{figure}[!t]
	\center{\includegraphics[width=0.7\textwidth]{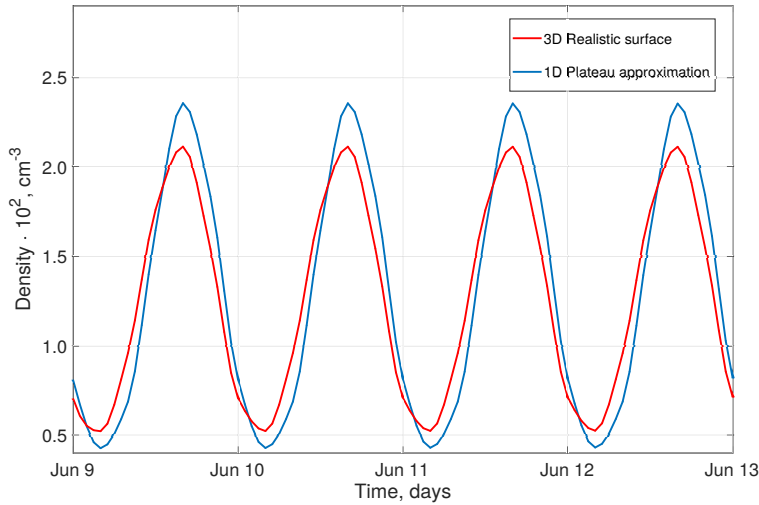}}
	\caption{The time-dependence of particle number density of SIMPs with $M = 1 \TeV$ and $\sigma = 60 \mb$ inside the DAMA detector around 
		June the 11th, obtained in the plateau approximation (blue curve) and for a realistic boundary surface (red curve).}
	\label{realsurf_vs_simple}
\end{figure}

\begin{figure}[!t]
	\begin{minipage}[h]{0.52\linewidth}
		\center{\includegraphics[width=1\textwidth]{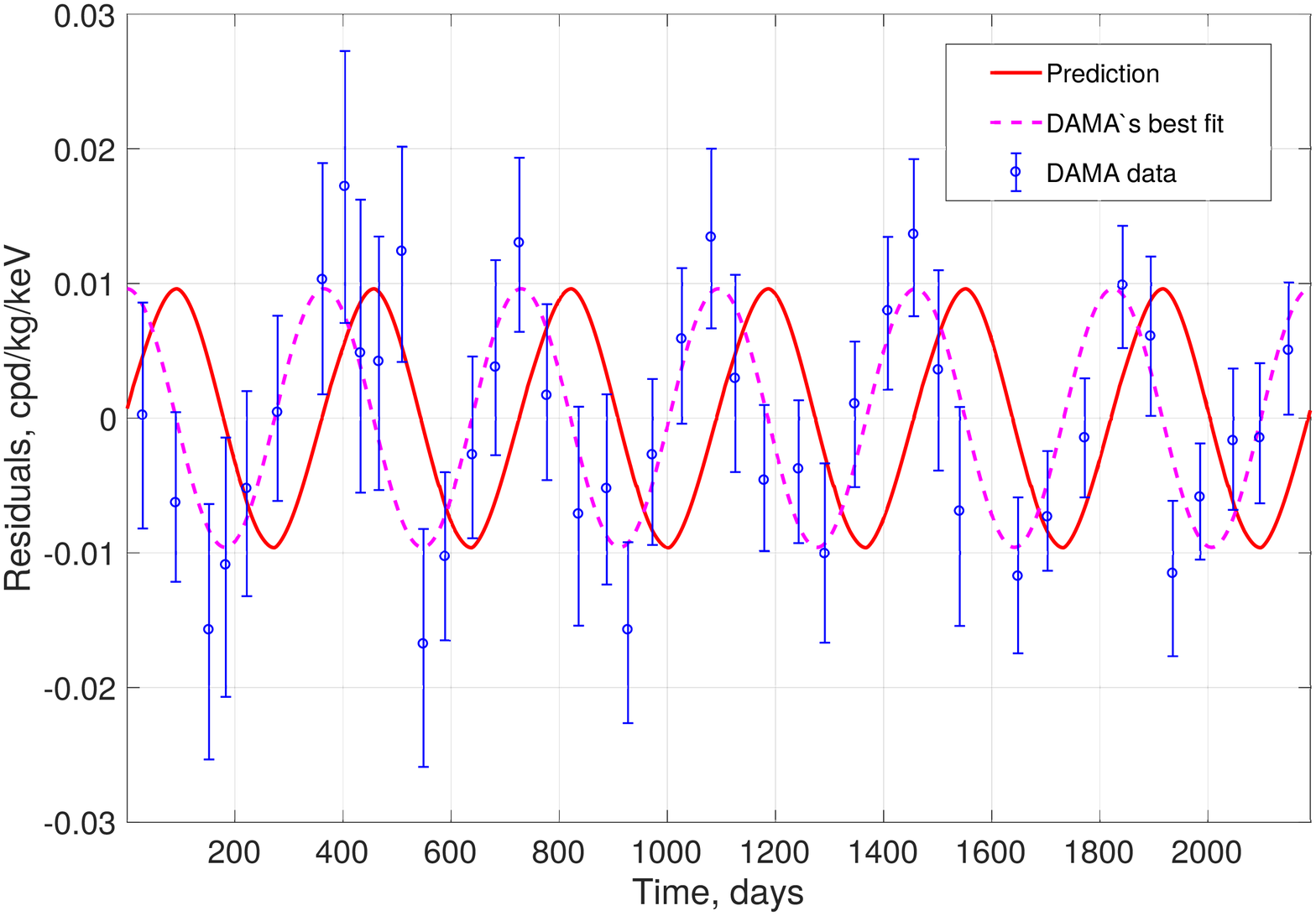}}
	\end{minipage}
	\begin{minipage}[h]{0.52\linewidth}
		\center{\includegraphics[width=1\textwidth]{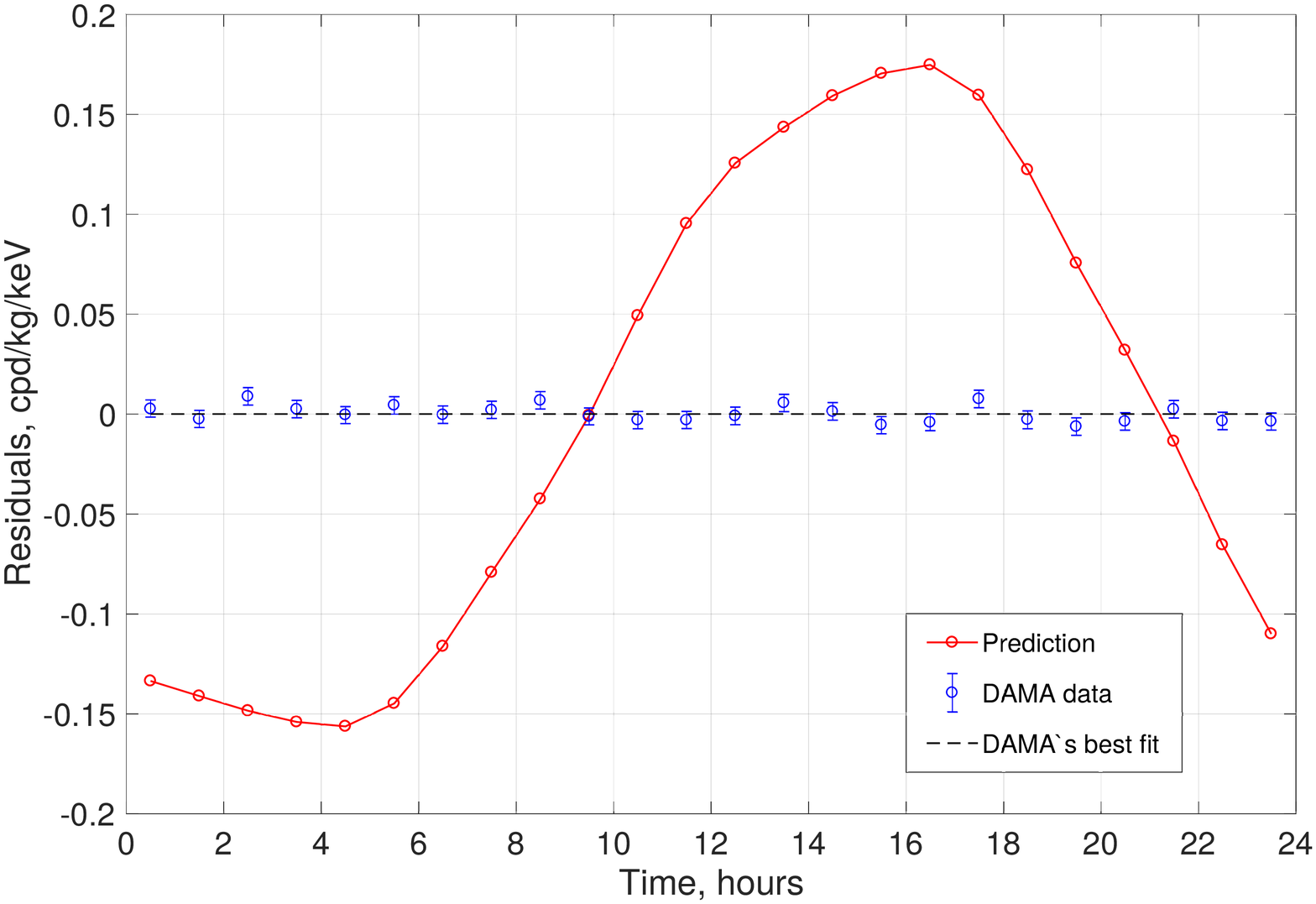}}
	\end{minipage}
	\caption{Annual (\textit{left}) and diurnal (\textit{right}) modulation of the residual rate of events in the DAMA detector in $2 - 6$ keV energy interval 
		for SIMPs with $M = 1 \TeV$ and $\sigma = 60 \mb$ calculated with the realistic boundary surface and compared to the experimental data.}
	\label{modulation_flat_mountain}
\end{figure}

Following the same routine as described above we compare our predictions to the DAMA data. The results we get for the realistic surface are not 
very different from those obtained in the plateau approximation (see Figs.~\ref{realsurf_vs_simple} and \ref{modulation_flat_mountain}). Thus, a 
more detailed approach doesn't solve the aforementioned problems with the DAMA explanation.

\section{Gravitational focusing}
\label{grav_focus}
Finally, we have considered the possibility that the annual phase is modified because the dark-matter flux must be bent by the Earth and the Sun (see e.g. \cite{Lee:2013wza,Herrero-Garcia:2018mky}). The magnitude of this effect is inversely proportional to the square of the velocity that an 
incoming particle has far away from the Sun and the maximum is expected around March. Since the maximum of the unperturbed annual dark matter velocity modulation is at the beginning of June, the effect of GF can modify the phase of the event rate in the direct-detection experiment 
like DAMA depending on the mass of dark-matter particles. In a WIMP scenario, this sets the bound on the velocity that a WIMP can have in order 
to create a recoil at the threshold energy. Unlike WIMPs, the signal from SIMPs is not due to recoils, so the rate of events is independent of their 
velocity and depends only on the total number density of dark matter particles, thermalised in the vicinity of the detector. In the cases considered 
here, the effect of gravitational focusing should alter the time variation of the thermalisation depth distribution (see Appendix \ref{append_distr}). 
We haven't taken this effect into account in our simulations, but if we assume that most of the particles thermalised at various depths have equal 
chances of getting inside the detector, than the time shift of the signal should be $\lesssim 20$ days \cite{Lee:2013wza}. However, a typical time 
delay of the annual modulation that we observe for SIMPs varies from $60$ to $200$ days, so even the maximal correction provided by 
gravitational focusing cannot improve the fit of the DAMA data sufficiently. Furthermore, the expected problematic diurnal modulation in case of 
SIMPs is not affected by the gravitational focusing of the Sun, although it can be influenced by that of the Earth. The implication of the latter effect 
for direct-detection searches was studied in \cite{Kouvaris:2015xga}. It appears that the Earth gravitational effect on the diurnal modulation of the 
total density of dark matter is comparable to the diurnal modulation coming from the rotation of the Earth w.r.t the Galactic frame and, thus, can 
hardly solve this issue for SIMPs. 

\section{Conclusion}
\label{conclusion}
In this paper, we have shown that the time dependence of the DAMA signal is incompatible with the hypothesis that it comes from
SIMPs. Of course, this does not rule out the possibility that SIMPs exist, but they cannot be the source of the signal that DAMA observes.
One has then to rely on models with very specific properties tuned to the data, if the signal is due to dark matter. 

Soon, this puzzle will 
have one more chapter as several experiments are under way or planned \cite{deSouza:2016fxg,DImperio:2018guh,Adhikari:2018ljm_new,Kim:2018wcl,Amare:2019jul,DiMarco:2019oda} using the same materials and techniques as DAMA does.
First results from COSINE-100 \cite{Adhikari:2018ljm_new} seem to suggest that the DAMA signal is not reproduced, and rule out spin-independent interactions as the cause. KIMS \cite{Kim:2018wcl} and NAIAD \cite{Alner:2005kt} only rule out part of the region of parameter space compatible with DAMA. The data from DM-Ice \cite{deSouza:2016fxg} and ANAIS \cite{Amare:2019jul} is consistent with a null-signal, though their statistics is insufficient to constrain the considered parameter region. Note that dark matter must go through the Earth to reach the South Pole, so a potential signal in DM-Ice could not be caused by SIMPs.
 
As these experiments are rather delicate, one should wait for more statistics and for more independent measurements to close this chapter.

\section*{Acknowledgements} 

We thank Igor P. Ivanov for pointing out the problem studied in this paper. % and Maxim Yu. Khlopov for discussions. 
This work was supported by the Fonds de la Recherche Scientifique - FNRS, Belgium, under grant No. 4.4501.15, and M.L. acknowledges the support of FRIA (F.R.S.-FNRS).

\appendix
\section{Solution of the drift-diffusion equation} \label{append_drift}

Let's consider a one-dimensional drift-diffusion equation in the form 

\begin{equation}
\frac{\partial v}{\partial t} = D \frac{\partial^2 v}{\partial z^2} + \beta\frac{\partial v}{\partial z}.
\end{equation}
This equation can be brought to the form

\begin{equation}
\frac{\partial u}{\partial t} = D \frac{\partial^2 u}{\partial z^2} ,
\end{equation}
by the following substitution

\begin{equation}
v = \exp(\mu z+\lambda t)u \quad \textrm{ with } \quad \mu = -\frac{\beta}{2D} \quad \textrm{ and } \quad \lambda = -\frac{\beta^2}{4D}.
\label{substit}
\end{equation}
Using this trick one can construct the Green's function $G'(z,\xi,t,\tau)$ for the operator 

\begin{equation}
L' = \frac{\partial }{\partial t} - D\frac{\partial^2 v}{\partial z^2}  - \beta\frac{\partial v}{\partial z} \, ,
\label{lprime_op}
\end{equation}
given the Green's function $G(z,\xi,t,\tau)$, which is a solution of the %following equation
operator

\begin{equation}
L = \frac{\partial}{\partial t} - D\frac{\partial^2}{\partial z^2}\, .
% - \beta\frac{\partial G}{\partial z} 
%= \delta(z-\xi)\delta(t-\tau) \, .
\end{equation}
Similarly to Eq.~\eqref{substit}, one has

\begin{equation}
G' = \exp(\mu(z-\xi) + \lambda(t-\tau)) G \, .
\end{equation}
Indeed, acting with the operator $L'$ %\ref{lprime_op}
on $G'$ and performing some simple derivations give

\begin{equation}
L'G' = \exp(\mu(z-\xi) + \lambda(t-\tau)) \, \delta(z-\xi)\delta(t-\tau)\, ,
\end{equation}
which is equal to $\delta(z-\xi)\delta(t-\tau)$ for any $z$ and $t$ and, thus, $G'$ is the Green's function for the operator $L'$. 

The solution of the drift-diffusion equation with the source $f(z,t)$ 
\begin{equation}
\frac{\partial v}{\partial t} = D \frac{\partial^2 v}{\partial z^2} + \beta\frac{\partial v}{\partial z} + f \, ,
\end{equation}
is then given by 

\begin{equation}
v(z,t) = \int d\tau \int d\xi \; \exp \left(-\frac{\beta(z-\xi)}{2D} - \frac{\beta^2 (t-\tau)}{4D}\right) \; G(z,\xi,t,\tau) \; f(\xi,\tau)\, ,
\end{equation}
where $G(z,\xi,t,\tau)$ is the Green's function of the simple diffusion equation without the drift term.

The one-dimensional case considered here can be easily generalised to multi-dimensional case.

\section{Derivation of the diffusion coefficient}
\label{diffusion_derivation}
%\todo[inline]{or drift velocity}

We are going to derive the diffusion coefficient for a gas of particles with mass $M$ spreading through a solid substance, which is comprised of $n$ molecules per unit volume with mass $m$. The latter particles are considered to be significantly lighter than the former and the cross section of their elastic scattering is $\sigma$. The gas is in thermal equilibrium with the surrounding matter at the temperature $T$.

\tikzset{
	big arrow/.style={
		decoration={markings,mark=at position 1 with {\arrow[scale=2,#1]{>}}},
		postaction={decorate},
		shorten >=0.4pt},
	big arrow/.default=black}

\begin{figure}[h!]
	\centering
	\begin{tikzpicture} [
	catg/.style={draw,dashed, minimum height=2em},
	thick/.style=      {line width=0.8pt}
	]
	
	%\draw [arrows={->[scale=2]}] (0,0) -- (8,0);
	\draw [big arrow] (0,0) -- (8,0);
	\draw [dashed](5,0.2) -- (5,1.8);
	\draw [dashed](5,-1.5) -- (5,-0.5);
	
	\draw (2,-0.5) -- (2,0.1);
	\draw (5,-0.5) -- (5,0.1);
	%\draw [arrows={<-[scale=2]}] (2,-0.4) -- (5,-0.4);
	\draw [big arrow] (5,-0.4) -- (2,-0.4);
	
	%\draw [arrows={->[scale=2]}] (2,0) -- (4,1);
	\draw [big arrow] (2,0) -- (4,1);
	\draw (2.7,0) arc[radius = 0.7, start angle= 0, end angle= 26];
	
	\node [text width=1cm, align=center] at (5.4, -0.4)  {$z_0$};
	\node [text width=1cm, align=center] at (8, -0.4)  {$z$};
	\node [text width=1cm, align=center] at (3.5, -0.7)  {$l$};
	\node [text width=1cm, align=center] at (3.1, 0.21)  {$\theta$};
	\node [text width=1cm, align=center] at (4, 1.3)  {$\vec{v}$};

	\end{tikzpicture}
	\caption{}
	\label{diagram}
\end{figure}

First, let us find the flux of particles $J^+$ crossing the plane $z_0$ (see Fig.~\ref{diagram}) in the direction of $z$. In the context of diffusion, the flux can be understood as the average number of particles $N^{+}$ per unit area, which pass through the given surface in between collisions, divided by the mean free time $\lr{\tau}$

\begin{equation}
J^+ =  \frac{\lr{N^{+}/A}}{\lr{\tau}} \, .
\label{J_pos}
\end{equation}
%where $n^{+}$ denotes the number density of particles moving in the direction of $z_0$ and $\tau$ is the time between scattering events. 
To derive a more useful expression, consider an infinitesimal layer of particles, located at the distance $l$ from the plane $z_0$. The number of particles per unit area of the layer $(l,l+dl)$, which move in the direction of z at angles ($\theta, \theta + d\theta)$ is given by

 \begin{equation}
\frac{1}{2}\, d\cos{\theta} \, dl  \; n(l) \, ,
%\exp\left(-\frac{l}{k\lambda\cos(\theta)}\right) 
\label{dee_en_plus}
\end{equation}
where the value of $\theta$ ranges from $0$ to $\pi/2$, because the particles moving in the opposite direction to $z_0$ can never reach 
it. % before the subsequent collision. 
However, not all of the particles determined by the above expression will manage to pass through $z_0$ -- some of them will experience a collision before. 
The distance $S$ that a particle travels before the next encounter is a random variable with an exponential probability density distribution
\begin{equation}
P(S) = \frac{1}{\lambda} \exp \left(-\frac{S}{\lambda}\,\right) ,
\label{PDF_lambda}
\end{equation}
where $\lambda = (n\sigma)^{-1}$ is the mean free path. %The corresponding time interval for a particle moving with the velocity $v$ is then $\tau = S/v$. 
Thus, the probability for particles defined by Eq.~\eqref{dee_en_plus} to reach the surface at the distance $l/\cos{\theta}$ is 
\begin{equation}
\int_{l/\cos{\theta}}^{\infty}dS \; \frac{1}{\lambda} \exp \left(-\frac{S}{\lambda}\,\right) = \exp \left(-\frac{l}{\lambda \cos{\theta}} \right).
\label{Probability_reach}
\end{equation}

However, we are interested not in the distance between collisions itself, but in the distance that a particle travels before significantly changing the direction of its velocity. When a heavy particle collides with a much lighter one, as in the case that we consider, it tends to kinematically preserve the direction of its movement (the effect known as the \textit{persistence of velocity}, see Chap. V in \cite{jeans1940introduction}).
Suppose, that the probability for a particle to proceed in the same direction after a collision is $q$. Then, the mean path length \underline{in one direction} is 
\begin{equation}
\lambda + q\lambda + q^2\lambda + \ldots = \lambda \, (1 + q + q^2 + \ldots) = \frac{\lambda}{1-q} = k\lambda \, ,
\label{mean_path_one_dir}
\end{equation}
where $k$ is a persistence coefficient, which we are going to derive later. Thus, for our purposes we should substitute this increased mean path value in Eq.~\eqref{PDF_lambda}.
%\begin{equation}
%P(S) = \frac{1}{k\lambda} \exp \left(-\frac{S}{k\lambda}\,\right) \, ,
%\label{PDF_lambda_k}
%\end{equation}

Now, gathering Eqs.~\eqref{dee_en_plus} and \eqref{Probability_reach}, we are ready to calculate the numerator in Eq.~\eqref{J_pos}

\begin{equation}
	\lr{N^{+}/A} = \frac{1}{2} \int_{0}^{1} d\cos{\theta} \int_{0}^{\infty} dl \exp \left(-\frac{l}{k\lambda \cos{\theta}} \right) n(l) \, . 
	\label{flux_numer}
\end{equation}

%Since the typical distances, at which we study the evolution of density in our problem, are much larger than $k\lambda$, we can approximately put
%Since the probability for a particle to reach $z_0$ decreases exponentially with the distance to it, 
We can expand the density in the vicinity of $z_0$ as 
 \begin{equation}
n(l) \approx n(z_0) -\frac{\partial n}{\partial z} \, l \, ,
\label{n_approx}
\end{equation}
where we have assumed that the density is decreasing in the direction of $z$, so that the net flux of particles through $z_0$ in that direction should be positive. Substituting this into Eq.~\eqref{flux_numer} and performing the integration one obtains
%Now, integrate over $\cos{\theta}$ and $l$ from $0$ to $\infty$ and, finally, obtain
%we are ready to calculate . For that we multiply Eq.~\eqref{dee_en_plus} by $v\tau = l/\cos{\theta}$, integrate it over $\cos{\theta}$ and $l$ from $0$ to $\infty$ and get 

\begin{equation}
\lr{N^+/A} = \frac{k\lambda}{2} \left(
\frac{n(z_0)}{2} - \frac{k\lambda}{3} \frac{\partial n}{\partial z}\right) .
\end{equation}
Dividing this by the mean free time 
\begin{equation}
\lr{\tau} = \int dS \, P(S) \int dv \, P(v) \, \frac{S}{v} = \frac{4\, k\lambda}{\pi \lr{v}}\, ,
\end{equation}
where $P(v)$ is the Maxwell-Boltzmann distribution for speeds and $\lr{v} = \sqrt{8kT/\pi M}$ is the mean speed of gas particles under consideration, we arrive at the expression for $J^+$

\begin{equation}
J^+ = \frac{\pi \lr{v}}{8} \left(\frac{n(z_0)}{2} - \frac{k\lambda}{3} \frac{\partial n}{\partial z}\right) .
\end{equation}

The flux $J^-$ crossing the plane $z_0$ in the opposite direction is given by the same formula, but the sign of the second term is positive due to the fact, that we redefine $l$ to be increasing in the same direction as $z$ and the sign in Eq.~\eqref{n_approx} changes. 
Then, the net flux $J$ through the surface $z_0$ is 
\begin{equation}
% J = J^+ - J^- = - \frac{k\lambda}{3} \sqrt{\frac{8kT}{\pi M}} \, \frac{\partial n}{\partial z} \, .
 J = J^+ - J^- = - \frac{\pi \, k\lambda \lr{v}}{12} \, \frac{\partial n}{\partial z} \, .
\end{equation}
%We assume that the density is decreasing in the direction of $z$, so that the flux in this direction is positive. 
Using Fick's first law of diffusion 
\begin{equation}
J = - D \, \frac{\partial n}{\partial z} \, ,
\end{equation}
we derive the expression for the diffusion coefficient
\begin{equation}
D  = \frac{\pi}{12} \, k\lambda \, \langle v \rangle \, .
\label{diffusion_coeff_express}
\end{equation}

Now, let us derive the persistence coefficient $k$. The probability $q$ of a particle to preserve the direction of its velocity after a random collision can be regarded as the value averaged over all collisions of the projection of the velocity after collision $\vec{v_f}$ onto the direction of its initial velocity $\vec{v_i}$, divided by the absolute value of the latter. For the particles that we consider, $\vec{v_f}$ can be derived from energy-momentum conservation
\begin{equation}
	\vec{v_f} = \frac{m}{M+m}\,  v_{\rm rel} \, \vec{n} + \frac{M\vec{v_i} + m\vec{u}}{M+m} \, ,
	\label{final_velocity}
\end{equation}  
where $\vec{u}$ is the velocity of a target particle, $v_{\rm rel} = |\vec{u} - \vec{v_i}|$ and $\vec{n}$ is a random unit vector. We shall first find the averaged projection over all possible directions of $\vec{u}$ and $\vec{n}$, while keeping the absolute values of $\vec{v}$ and $\vec{u}$ fixed. Clearly, the first term in \eqref{final_velocity} averages to $0$, so that the expectation of the component of velocity of either particle after collision in any direction is equal to the component of $V_{\rm CM}$ in that direction. We should not, though, suppose that all directions of $\vec{u}$ are equally likely, because the probability of collision with any two velocities is proportional to $v_{\rm rel}$. Thus the value of the component of $\vec{v_f}$ in the direction of $\vec{v_i}$ averaged over all possible kinematic directions is 

\begin{equation}
	\left\langle \vec{v_f} \cdot \frac{\vec{v_i}}{v_i} \right\rangle = \frac{1}{\int_{-1}^{1} d\cos{\omega} \,  v_{\rm rel}} \int_{-1}^{1} d\cos{\omega} \; \frac{\left(M v_i + m u \cos{\omega}\right) v_{\rm rel}} {m+M} \, ,
\end{equation}
where $\omega$ is the angle between $\vec{v_i}$ and $\vec{u}$. %Dividing this expression by $v_i$ %and changing variables $\cos{\omega} \rightarrow v_{\rm rel}$
%performing the integration over $v_{\rm rel}$ 
%one obtains the expression for $q$ as a function of speeds 
Then, the probability $q$ can be expressed as a function of speeds

\begin{equation}
	q(u,v_i) = \frac{1}{v_i}\left\langle \vec{v_f} \cdot \frac{\vec{v_i}}{v_i} \right\rangle = \frac{M}{m+M} + \frac{u^2 + v_i^2}{2v_i^2} \frac{m}{m+M}  - \frac{1}{2 v_i^2} \frac{\int dv_{\rm rel} \; v_{\rm rel}^4}{\, \int dv_{\rm rel} \; v_{\rm rel}^2} \, \frac{m}{m+M} \, ,
\end{equation}
where $v_{\rm rel}$ varies from $|v_i-u|$ to $v_i+u$. Since we assume that $M \gg m$, it is reasonable then to assume also that $u > v_i$ most of the time. Performing the integration and combining the second and the third terms, we get

\begin{equation}
q(u,v_i) = \frac{M}{m+M} +  \frac{v^2 - 5u^2}{5(v^2+3u^2)} \frac{m}{m+M} \, .
\end{equation}
Taking into account again, that $M \gg m$, we can suppose that on average $u \gg v$ and hence
\begin{equation}
q \approx \frac{M}{m+M} - \frac{1}{3} \frac{m}{m+M} \, .
\label{q_express}
\end{equation}
We see, indeed, that the probability for a massive particle to follow the same direction after collision with a much lighter particle is very close to $1$, so the mean free path length in one direction is extended. Finally, from Eq.~\eqref{mean_path_one_dir} and Eq.~\eqref{q_express} we can write down the persistence coefficient $k$

\begin{equation}
k = \frac{3}{2} \frac{m+M}{m} = \frac{3}{2} \frac{M}{\mu} \, ,
\end{equation}
where $\mu$ is the reduced mass. At last, plugging this result back into Eq.~\eqref{diffusion_coeff_express} one obtains the expression for the diffusion coefficient

\begin{equation}
D  = \frac{\pi}{8} \, \frac{M}{\mu} \lambda \langle v \rangle \, .
\end{equation}

Interestingly enough, there is another way to obtain this expression. One should consider the gas of heavy particles in a force-field, derive its mean (drift) velocity in this field by averaging over all possible directions in a random collision and apply the Einstein relation. %(see text above Eq.~\eqref{drift_vel}).

\section{Dark-matter velocity and thermalisation depth distributions} \label{append_distr}

We assume that SIMPs follow the same velocity distribution as the rest of dark matter. For the dark-matter velocity distribution in the halo rest frame we take a bounded Maxwellian distribution

\begin{equation}
\omega_v(v) = \exp \left(-\frac{v^2}{v_{0}^2}\right) \, \theta \left( v_{\rm esc} - v \right),
\end{equation}
where $\theta$ denotes the Heaviside step-function, $v_{0}$ is the dispersion velocity of the halo and $v_{\rm esc}$ as the Galactic escape velocity at the position of the Sun. We adopt the following values of these parameters $v_{0} = 220 \km / \s$ \cite{1986MNRAS_221_1023K} and $v_{\rm esc} = 544 \km / \s$ \cite{Lavalle:2014rsa}. Here we omit the normalisation factor of the distribution for simplicity. In the laboratory frame dark matter velocity distribution is different

\begin{equation}
\omega_v(\vec{v},t) = \exp \left(-\frac{(\vec{v} - \vec{v}_{\rm lab}(t))^2}{v_{0}^2}\right) \, \theta \left( v_{\rm esc} - \left| \vec{v} - \vec{v}_{\rm lab}(t) \right| \right),
\end{equation}
where $\vec{v}_{\rm lab}$ is the lab velocity in the Galactic frame. This velocity is a periodical function of time and depends on the location on the globe. For our analysis we calculate $\vec{v}_{\rm lab}$ at the location of LNGS following the procedure described in \cite{Bernabei:2014jnz}.

\begin{figure}[h!]
	\center{\includegraphics[width=0.8\textwidth]{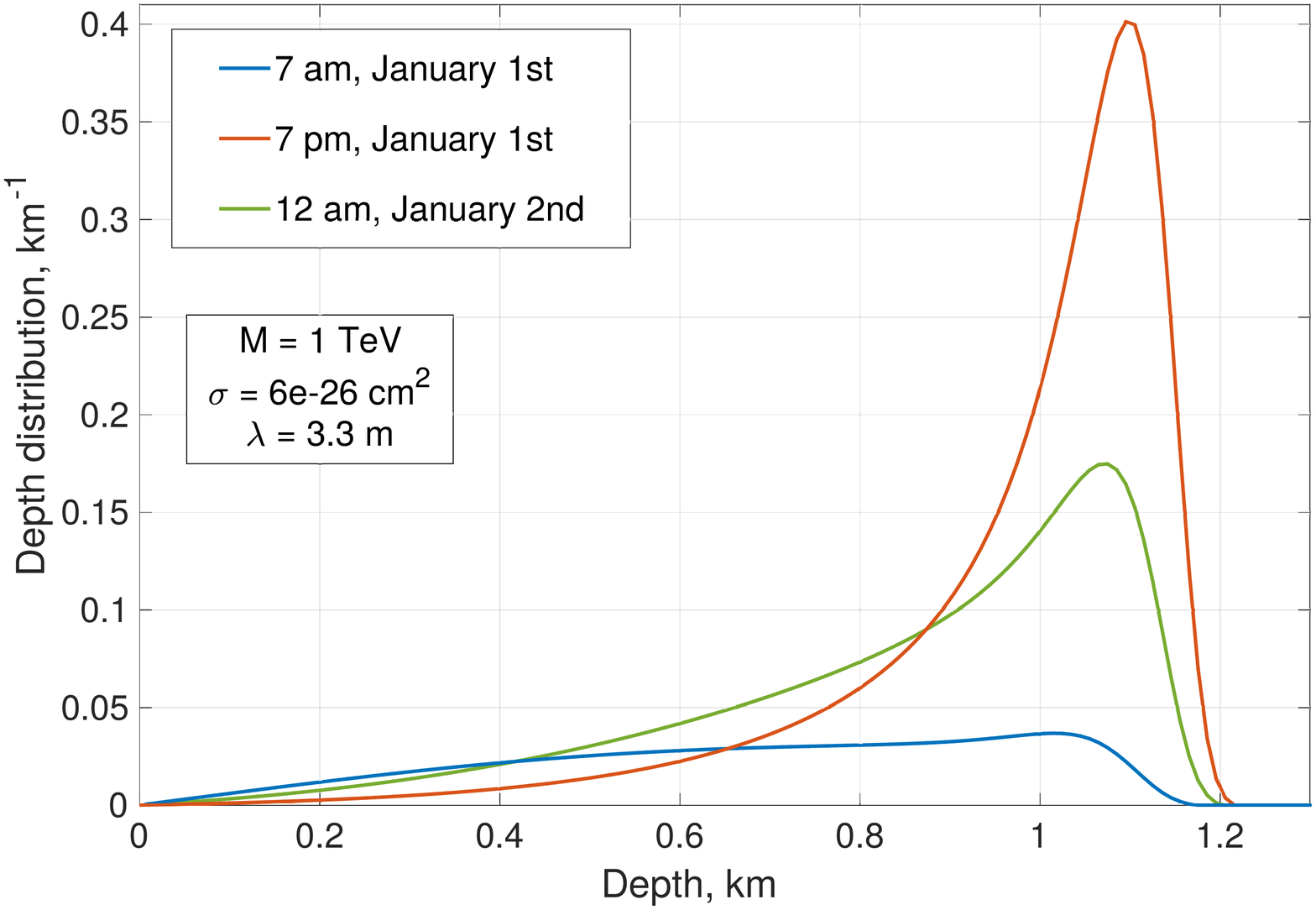}}
	\caption{Exemplative distributions of the density of dark matter particles at the depth at which they reach thermal velocities. }
	\label{modulation_flat_realistic}
	\label{depth}
\end{figure}

Using the relation \eqref{l_th} between the velocity $v$ of the dark-matter particle and the distance $l$ it travels in the ground before thermalisation one can transform the velocity distribution $\omega_v(\vec{v},t)$ into the thermalisation depth distribution $\omega_l(\vec{l},t)$. Here $\vec{l}$ denotes the vector which connects the point where dark-matter particle enters the ground and the point where it thermalises. The direction of $\vec{l}$ obviously coincides with the direction of $\vec{v}$, so the components of these vectors are simply related as $l_i = (l/v)v_i$. Following the probability density equality 

\begin{equation}
\omega_v(\vec{v},t)\, d^3\vec{v} = \omega_v\left(\frac{v}{l}\vec{l},t\right) J(l) \, d^3 \vec{l}\, ,
\end{equation}
where 
\begin{equation}
J(l) = \frac{mv^3(l)}{M\lambda l^2}
\end{equation}
is the Jacobian of the transformation from the velocity space to the distance space. Thus, for the thermalisation depth distribution one obtains

\begin{equation}
\omega_l(\vec{l},t) = \frac{mv^3(l)}{M\lambda l^2} \, \exp \left(-\frac{(\vec{v}(l) - \vec{v}_{\rm lab}(t))^2}{v_{0}^2}\right) \, \theta \left( v_{\rm esc} - \left| \vec{v}(l) - \vec{v}_{\rm lab}(t) \right| \right).
\end{equation}
This leads to distributions illustrated in Fig. \ref{depth}.

% Bibliography

%\bibliographystyle{JHEP}
%\bibliography{References}
	
\providecommand{\href}[2]{#2}\begingroup\raggedright\endgroup

\end{document}